\documentclass[aps,prl,twocolumn,showpacs,preprintnumbers,amsmath,amssymb,superscriptaddress]{revtex4}
\usepackage{graphicx, epsfig}% Include figure files
\usepackage{dcolumn}% Align table columns on decimal point
\usepackage{bm}% bold math
\usepackage{color}

\newcommand{\cmm}[0]{cm$^{-1}$ }

\newcommand{\etal}{\emph{et al.~}}

\begin{document}

%\preprint{PRL/???}

\title{Charge carrier interaction with a purely electronic
collective mode:  Plasmarons and the infrared response of
elemental bismuth}

\author{Riccardo Tediosi}
%\email{riccardo.tediosi@physics.unige.ch}
\affiliation{D\'{e}partement de Physique de la Mati\`{e}re
Condens\'{e}e, Universit\'{e} de Gen\`{e}ve, quai Ernest-Ansermet
24, CH1211 Gen\`{e}ve 4, Switzerland.}
\author{N.~P.~Armitage}
\affiliation{D\'{e}partement de Physique de la Mati\`{e}re
Condens\'{e}e, Universit\'{e} de Gen\`{e}ve, quai Ernest-Ansermet
24, CH1211 Gen\`{e}ve 4, Switzerland.} \affiliation{Department of
Physics and Astronomy, The Johns Hopkins University, Baltimore, MD
21218, USA.}
\author{E. Giannini}
\author{D. van der Marel}
\affiliation{D\'{e}partement de Physique de la Mati\`{e}re
Condens\'{e}e, Universit\'{e} de Gen\`{e}ve, quai Ernest-Ansermet
24, CH1211 Gen\`{e}ve 4, Switzerland.}

\date{\today}% It is always \today, today,

\begin{abstract}

We present a detailed optical study of single crystal bismuth using
infrared reflectivity and ellipsometry.  Colossal changes in the
plasmon frequency are observed as a function of temperature due to
charge transfer between hole and electron Fermi pockets. In the
optical conductivity, an anomalous temperature dependent
mid-infrared absorption feature is observed.  An extended Drude
model analysis reveals that it can be connected to a sharp upturn in
the scattering rate, the frequency of which exactly tracks the
temperature dependent plasmon frequency.  We interpret this
absorption and increased scattering as the first direct optical
evidence for a charge carrier interaction with a collective mode of
purely electronic origin; here electron-plasmon scattering. The
observation of a \emph{plasmaron} as such is made possible only by
the unique coincidence of various energy scales and exceptional
properties of semi-metal bismuth.
\end{abstract}

\maketitle

Elemental semi-metals, such as graphite and bismuth, are materials
of much long term interest due to their exceptional properties,
including large magnetoresistive and pressure dependent effects
\cite{brandt,balla65,edelman76}. In the case of bismuth these
properties derive from its low carrier number, ($\approx 10 ^{-5}$
electrons per atom), reduced effective masses ($\approx 10 ^{-2}$
electron masses), small Fermi wavevector ($\approx 40$ nm), long
mean free path ($\approx 1$ mm), and large high frequency
dielectric constant ($\epsilon_\infty \approx 100$).

A number of recent results are causing an increased interest in
these materials, both from the side of fundamental solid-state
physics as well as applications potential. For instance a field
dependent crossover reminiscent of the 2D metal/insulator transition
in MOSFETs \cite{kravchenko} has been observed in both graphite and
bismuth \cite{hebard}.  Isolated single layers of graphene have been
shown to have novel transport properties and an anomalous
quantization of the Quantum Hall effect resulting from their low
carrier number and exceptional zero mass Dirac cone dispersion
relation \cite{graphene1,graphene2}. Moreover there continues to be
interest in bismuth for studies of quantum confinement
\cite{quantumconfinement}. On the technical side, advances in film
growth \cite{chien}, anomalously long spin diffusion lengths and
very large magnetoresistive response makes bismuth useful for
possible incorporation in nanomagnetometers, magnetooptical devices
and spintronics applications
\cite{chong,boero,grande,leespindiffusion}.

In principle transport phenomena in bismuth should be well
described by the conventional theory of metals, but due to its
exceptional parameters, there are substantial departures from
standard metallic behavior. For instance, electronic energy
scales, like the Fermi energy, are very low giving strongly
temperature dependent effective masses and charge densities.
Moreover, the material's very low carrier density opens up the
possibility - at least in principle - for novel plasmonic effects
and strong electron-electron interactions due to the relative
scales between potential and kinetic energy at low charge
densities \cite{mahan}.

Despite the scientific and technological interest in bismuth, its
optical and infrared properties have been under-investigated.  In
this letter we present detailed temperature dependent optical
measurements over the full optical range from FIR to UV of single
crystal bismuth. We observe a narrow Drude peak which has a plasma
frequency value consistent with the low carrier number. Colossal
changes in the plasma frequency are observed as a function of
temperature, due to charge transfer between electron and hole
pockets.  We find an anomalous mid-infrared absorption in the real
part of the conductivity.  An extended Drude model analysis reveals
that the scattering rate has an abrupt onset at a temperature
dependent energy scale which is found to be almost exactly
coincident with the independently measured plasmon energy. This is
the first direct optical observation of a strongly coupled
electron-plasmon elementary excitation - a $plasmaron$.

\begin{figure}
\includegraphics[bb= 0 0 453 425,clip,width=8cm]{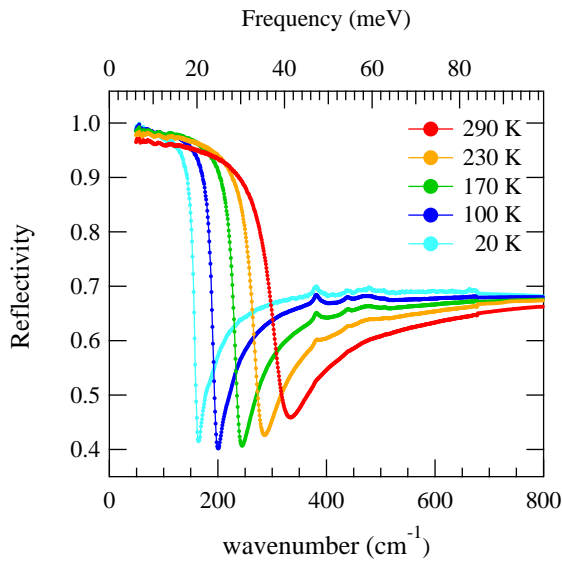}
\caption{\label{fig:reflectivity} (color online) Single-crystal
bismuth reflectivity vs wavenumber ($\omega/2\pi c$) for
temperature T = 290 (red), 230 (orange), 170 (green), 100 (blue)
and 20 K (cyan). The pronounced shifting of the minimum in the
range $180-350$ \cmm represents the position of the screened
plasma frequency $\omega_{p}^{*}$. The small peak around 380 \cmm
is an experimental artifact due to a small nonlinearity in a
strong absorption of the interferometer's beam splitter.}
\end{figure}

Single crystal bismuth was grown by a modified
Bridgman-Stockbarger technique in a vertical three-zone furnace. A
silica tube was filled with $\sim 5$ grams of 99.999\% pure Bi
powder (Cerac) and sealed under vacuum. The ampoule was held
vertically in a three-zone furnace and annealed above the melting
point of Bi ($T_m$ = 271.4 $^\circ$C) for 10 hours before
decreasing the temperature at a rate of 30 $^\circ$C/h, while
keeping a temperature gradient of 10-15 $^\circ$C/cm.  The
crystals were cleaved from the as-grown boule along a plane
perpendicular to the trigonal direction [001] at LN$_2$
temperatures.  X-ray powder diffraction revealed that the
mirror-like cleavage surfaces were [110] planes perpendicular to
the trigonal axis which were subsequently used as reflecting
surfaces for optical experiments.

We measured the DC resistivity and optical spectra in the frequency
range from 50 \cmm (6.2 meV) to 30000 \cmm (3.8 eV) combining
infrared (IR) reflectivity via FT spectroscopy and ellipsometry in
the VIS-UV energy range. In the IR experiment the sample was mounted
in a quasi-normal incidence configuration
($\theta_{\textrm{inc}}=11^\circ$) and reflected signal intensity
was recorded during slow temperature scans in the temperature range
from 290 K down to 20 K with a resolution of approximately 1 K. The
absolute value of the reflectivity $R(\omega, T)$ was calculated
using a reference gold layer evaporated \emph{in situ} on the sample
surface.

In Fig.~\ref{fig:reflectivity} the reflectivity is presented for
five selected temperatures in the far-infrared spectral range. A
spectacular shift of the reflectivity edge from around 333 \cmm at
room temperature to a value of 164 \cmm at 20 K indicates a strong
reduction of the plasmon frequency with cooling.  This derives from
a change in density due to thermal charge transfer from electron to
hole pockets. The small feature present at 380 \cmm at all
temperatures is a non-linear effect of the detector as a result of
absorption in the beamsplitter and is not an intrinsic feature of
bismuth.  The ellipsometry and IR data were combined using a
Kramers-Kronig consistent variational fitting procedure
\cite{alexey}. This allows the extraction of all the significant
frequency and temperature dependent optical properties like for
instance, the complex conductivity $\hat{\sigma}(\omega, T)=\sigma_1
+ i\sigma_2$. Note that although it does not affect our conclusions
either way, the 380 \cmm artifact has been removed from the data
used to generate the plots in subsequent figures. The extended
frequency range and ellipsometry used in our experiment allows us to
extract more accurate parameter values from the subsequent analysis
than what has been reported in previous work \cite{boyle60} where
just the FIR spectral range was analyzed.

\begin{figure}
\includegraphics[bb= 0 0 453 396,keepaspectratio,width=8cm]{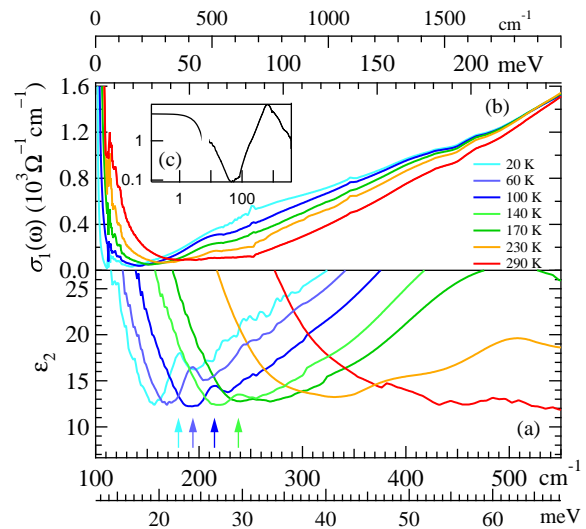}
\caption{\label{fig:sigma} (color online) (a) Optical conductivity
derived from Kramer-Kronig analysis for selected temperatures.
 (b) The imaginary dielectric constant $\epsilon_2$ over a much
 smaller energy range that emphasizes the remarkable prepeak
 structure. (c) Log-Log scale plot of $\sigma_1(\omega)$ for $T=290$ K, units are meV. }
\end{figure}

In Fig.~\ref{fig:sigma}(a) the optical conductivity
$\sigma_1(\omega)$ is presented for corresponding data of
Fig.~\ref{fig:reflectivity} in the FIR-MIR range. The plot reveals
two main features developing with decreasing temperature; the first
is a progressive narrowing of the Drude peak from an
half-width-half-maximum value of 42.3 \cmm at room temperature down
to a value of 3.3 \cmm at 20 K.  Additionally, we observe a dramatic
appearance and strengthening of an absorption centered around 700
\cmm and characterized by an onset approximately around 350 \cmm at
room temperature which appears to shift downward as the temperature
is lowered.  In the inset to Fig~\ref{fig:sigma} the conductivity is
presented over the full measurement range at room temperature;  a
prominent peak is evident in the figure centered around 5500 \cmm
whose energy and temperature dependence are compatible with the
direct interband transition at the \emph{L}-symmetry point.

While the narrowing of the Drude peak is typical behavior for a
metal whose DC conductivity increases at lower temperatures as a
consequence of the reduction of the scattering processes, the
appearance of an MIR absorption is unusual. In fact, a closer look
shows an even more interesting aspect, as shown in Fig. 2b where we
plot a greatly expanded view of the related quantity, the imaginary
dielectric constant $\epsilon_2 = 4\pi \sigma_1 /\omega$; the low
energy onset of this MIR absorption feature is preceded by a small,
but distinct and temperature dependent \emph{prepeak} absorption
structure.

In a previous study, the MIR absorption has been assigned to the
threshold for direct interband transitions at the $L$ point
\cite{boyle60}. Certainly interband transitions play a role in part
of this energy range, but the energy scale of the onset and prepeak
are not quite right for them to be the entire contribution. The
interband gap reported at low temperature is $13.7$ meV
\cite{edelman76} and combined with an $L$ point Fermi energy E$_F$,
gives a minimum threshold for direct absorption of $E_c + 2 E_F$ of
$67.1$ meV (540 \cmm), which is much bigger than the onset. The
discrepancy even increases at low temperature where the onset falls
to 125 \cmm. It is possible that the region of the absorption onset
is partially affected by an indirect interband phonon coupled
process, as recently proposed for a similar absorption in bismuth
nanowires \cite{black2003}. Although this process has been
investigated theoretically \cite{stepanov04,tediosi2007} we observe
that our experimental data deviate considerably from the
expectations, in particular in the region where the prepeak appears.
In the rest of this work we will thus concentrate on the explanation
that such anomaly derives from an electron-plasmon interaction.

\begin{figure}
\includegraphics[bb= 0 0 453 396, keepaspectratio,width=8cm]{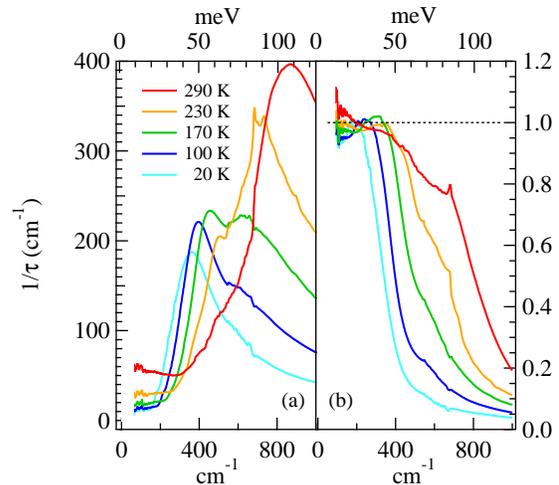}
\caption{\label{fig:tau_mass} (color online) Extended-Drude analysis
results: (a) the low frequency scattering rate $\tau^{-1}(\omega)$
progressive falls as the temperature is lowered.  An approximately
frequency independent region is interrupted by a sharp onset in
scattering.  (b) The effective mass $m^*(\omega)/m$ is flat as over
a similar range. $m$ is defined as the band-mass of the carrier for
the specific crystal orientation used.}
\end{figure}

In Fig 3, we analyze the complex conductivity data in terms of an
extended Drude model.  We extract the frequency dependent
scattering rates $\tau^{-1}(\omega)$ and effective masses
$m^*(\omega)$ via the relations
\begin{eqnarray}\label{eq:tau_mass}
\nonumber
  \tau^{-1} (\omega) &=& -(\omega_p^2/\omega) \textrm{Im} \left( \varepsilon(\omega) - \tilde{\varepsilon}_\infty
  \right)^{-1}\\
  m^*(\omega)/m_e &=& -(\omega_p^2/\omega ^2) \textrm{Re} \left( \varepsilon(\omega) - \tilde{\varepsilon}_\infty
  \right)^{-1}
\end{eqnarray}
where $\omega_p$ is the plasma frequency and
$\tilde{\varepsilon}_\infty$ represents the temperature dependent
interband contribution to the dielectric constant.  We should stress
here the fact that the extended Drude model is strictly valid only
in the region where interband transitions do not play a major role.
We see in Fig.~\ref{fig:tau_mass} that the data is relatively well
described at the lowest frequencies within the usual Drude framework
where the scattering rates and masses are frequency independent over
roughly the same frequency interval. However we observe a sharp
onset in the scattering rate at a well-defined temperature dependent
energy scale. This reflects the MIR absorption pointed out
previously and makes $\tau^{-1}(\omega)$ deviate from the constant
value expected from a simple Drude model.

\begin{figure}
\includegraphics[bb= 0 0 453 396, keepaspectratio,clip,width=8cm]{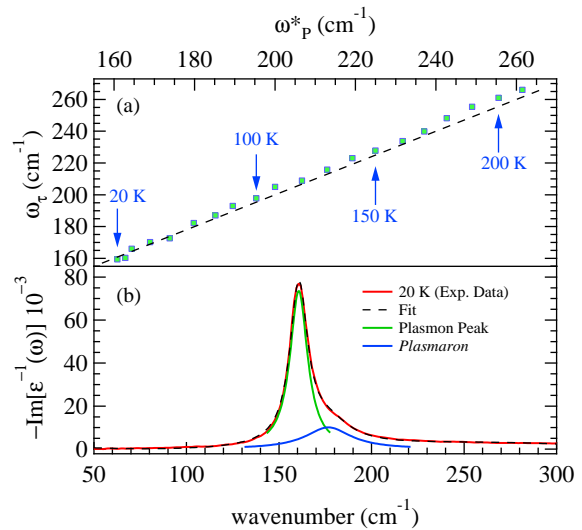}
\caption{\label{fig:scattering_onset} (color online) (a) A
parametric plot $\omega_\tau$ vs.~$\omega_p^*$ shows a slope of 1
supporting the hypothesis of an electron-plasmon interaction. (b)
The 20K electron energy loss (EEL) function (red) presents the
main plasmon peak and a \emph{plasmaron} peak appearing as a
shoulder of the main one. }
\end{figure}

As noted earlier, according to band structure parameters
\cite{vecchi74} the position of the scattering onset $\omega_\tau$
is too high to derive from a direct interband process.  In fact,
the temperature dependence of this absorption's onset and prepeak
is reminiscent of the large temperature dependence of the plasma
frequency itself. We observe that the onset almost exactly tracks
the independently measured plasma frequency as a function of
temperature, as shown in the parametric plot
Fig.~\ref{fig:scattering_onset}(a) where we plot the frequency of
this onset or ``kink'' in the $\varepsilon_2(\omega)$ function
$\omega_\tau$ vs.~$\omega_p^*$ as defined by the zero crossing of
the experimental $\varepsilon_1(\omega)$ and the onset is defined
by the energy position of the prepeak's half maximum on its low
frequency side. Based on this parametric plot in
Fig.~\ref{fig:scattering_onset} we can conclusively identify the
plasma frequency as setting the energy scale for the observed
absorption process. The plasmon energy changes by almost a factor
of two over almost the entire frequency range, but continues to
set the scale for the increased scattering throughout.

Due to its longitudinal character, direct excitation of a plasmon by
an incident electromagnetic wave is generally not possible. However,
a number of scenarios may exist to induce an electron-plasmon
coupling effect. Previously, an explanation for the appearance of
the MIR peak has been given in terms of an impurity-mediated
electron-plasmon coupling \cite{gerlach74, gerlach76, classen86,
mycielski78}.  It was proposed that enhanced electron-charged
impurity scattering is found near $\omega_p$ due to the divergence
of $1/\epsilon$. Although our data are in qualitative agreement with
such a scenario, the model used by Gerlach \etal \cite{gerlach76}
has a quantitative agreement only by considering an exceptionally
large charged impurity concentration ($N = 1.5\times 10^{19}$
cm$^{-3}$). This is approximately two orders of magnitude greater
than the carrier concentration itself and implies a \emph{charged}
impurity concentration of 1 part in $10^4$, which we consider to be
unrealistic considering the high purity of our samples.

In contrast, we propose that we are observing the excitation of
plasmons via a decay channel of the excited electron-hole pairs.
This interaction is essentially identical to that considered in the
context of electron - phonon \cite{marsiglio01} or electron - magnon
interactions \cite{carbotte99}.  Such an interaction may in fact be
captured within the same Holstein Hamiltonian that is used to
describe the electron - longitudinal phonon coupling to treat
polarons and so this collective excitation has been called a
$plasmaron$ \cite{lundqvist1,lundqvist2} in theoretical treatments.
Such an excitation is only possible optically in a system where
translational symmetry has been broken first by, for instance,
Umklapp scattering or disorder that moves oscillator strength to a
frequency region near the plasmon energy. In normal metals such an
interaction is completely unobservable as the plasmon energy scales
are many orders of magnitude higher than transport ones. This is, to
the best of our knowledge, the first unambiguous observation of
charge carrier scattering with a collective bosonic mode of purely
electronic origin.

The possibility also exists that we are observing not an
electron-hole decay channel, but instead a direct 3 body excitation
of an electron-hole pair and a plasmon with a net momentum $q \sim
0$. Such processes are possible by going beyond the usual RPA and
considering electron-electron interactions (electron-plasmon in the
present case) mediated by the crystal potential \cite{mahan}. In the
case of a translational invariant system, the standard Landau-Fermi
liquid treatment holds since the photon-induced electron-hole pair
are momentum conserving processes. Introducing symmetry breaking
terms, like strong Umklapp scattering or disorder, makes such 3 body
processes possible.   Experimentally this effect may be directly
observable using optical spectroscopy since the longitudinal
collective mode would be partially coupled to a transverse one
\cite{turlakov2003}.

Within the electron-hole plasmon decay channel scenario one might
try to model the scattering rate using an expression for the plasmon
density of states $D(\omega)$ and a simple scheme for coupling of
the spectrum to electronic excitations \cite{carbotte99,
marsiglio01}. Unfortunately, although aspects of our data are
qualitatively consistent with such a scenario, such a calculation is
hard to compare with the data exactly, due to the onset of the
interband contribution.  The low energy prepeak structure is
interesting and is $not$ captured within any simple models. We
speculate that it is a resonant effect due to enhanced electronic
interaction near $\omega_p$ deriving from the divergence of
$1/\epsilon$.

In view of these difficulties, we show in
Fig.~\ref{fig:scattering_onset}(b) the 20 K electron energy loss
function $\textrm{EEL}(\omega) = \textrm{Im}\left\{
-\varepsilon^{-1}(\omega) \right\}$ which should be relatively
insensitive to the contributions of interband terms. The plot
presents a prominent peak centered at $\omega_p^*$ but also an high
frequency shoulder. The total EEL function can be decomposed using
two lorentzian oscillators; one centered at the screened plasma
frequency $\omega_p^* = 160.7$ \cmm with a width $\gamma_1 = 5.74$
\cmm and another one at $\omega_2 = 176.4$ \cmm with $\gamma_2 =
15.6$ \cmm. The position of this latter peak corresponds exactly to
the position of the absorption feature seen in
$\varepsilon_2(\omega)$ thus demonstrating its admixture of
longitudinal plasmonic character.

In conclusion we have made the first optical observation of an
electron-plasmon interaction  - a plasmaron.  This observation is
made possible only by the low carrier density in Bi, a very large
$\epsilon_\infty$, and a lack of optically active phonons.  This
work raises various questions about how these renormalization
effects feed back on the low energy properties of this material.
It is possible that reducing the charge density further, perhaps
through the application of pressure, may enhance such interactions
and drive the system into an anomalous metallic state. Pressure
dependent optical studies may prove to be a useful probe in this
regard.  It would also be interesting to search for similar
effects in other semi-metals like graphite or single-layer
graphene.

The authors would like to thank M. Dressel, H.D. Drew and A.J.
Millis for various illuminating conversations.  The work at the
University of Geneva is supported by the Swiss National Science
Foundation through the National Center of Competence in Research
``MaNEP''.  NPA has been also supported via the NSF's
International Research Fellows program.

\bibliography{BismuthBib_etal}%

\end{document}